# The puzzle of magnetic resonance effect on the magnetic compass of migratory birds

*K.V.Kavokin*

A.F.Ioffe Physico-Technical Institute
Polytechnicheskaya 26, St.-Petersburg 194021, Russia
e-mail: kidd.orient@mail.ioffe.ru

Experiments on the effect of radio-frequency (RF) magnetic fields on the magnetic compass orientation of migratory birds are analyzed using the theory of magnetic resonance. The results of these experiments were earlier interpreted within the radical-pair model of magnetoreception. However, the consistent analysis shows that the amplitudes of the RF fields used are far too small to noticeably influence electron spins in organic radicals. Other possible agents that could mediate the birds' response to the RF fields are discussed, but apparently no known physical system can be responsible for this effect.

*1. Introduction.*

Hundreds of publications are devoted to the hypothetic magnetic sense of animals (for a review, see [Wiltschko and Wiltschko 1995, Wiltschko and Wiltschko 1996, Wiltschko and Wiltschko 2005, Johnsen and Lohmann 2005]), yet the phenomenon remains elusive and is often regarded with skepticism. The reason is that no direct physiological evidence of magnetoreception has been obtained so far. As we humans are apparently deprived of this sense, and other animals cannot tell us whether they feel magnetic fields or not, nearly all the available information has been obtained from behavioral experiments. These experiments are rarely quite unambiguous and normally require outstanding skills and experience in working with the animal species in question, to exclude the influence of innumerable interfering factors. No wonder that, in spite of the abundance of published experimental results, no factual base has been built for constructing a conclusive physical theory of magnetoreception.

However, some results, obtained lately in experiments with migratory birds, allow quantitative interpretation and might become the basis for such a theory.

Birds are certainly a "privileged" class of animals in what concerns the studies of the magnetic sense, for two main reasons. Firstly, the ability of migratory birds to find their way over great distances is known to the mankind for ages; the analogy with the methods of human navigators brought about the idea of birds' using a magnetic compass as long ago as in $19^{th}$ century[1]. Second, and more important for the modern science, factor is the genetically prescribed *circannual rhythm* of the bird's life, that gives a great advantage to the experimentalist. In certain periods of the year, when the bird in nature should migrate, its physiology and behavior dramatically change [Berthold 1996]. In particular, the bird, even kept in a cage, demonstrates increased motional activity. Very often this *migratory restlessness* is accompanied by attempts to move in the direction corresponding to the seasonal migration route. The preferential direction of movements of caged birds, recorded by various technical means, was shown to be determined from celestial cues like the Sun or stars (including artificial ones, demonstrated to the bird in a planetarium) [Emlen 1967a, Emlen 1967b]. Many bird species, when deprived of the possibility to take bearing from any visible object, still show seasonally appropriate direction of movements. This direction was reported to change when the laboratory magnetic field was deflected with electromagnets

---

[1] According to [Wiltschko and Wiltschko 1996], by A. von Middendorff (1859) and C.Viguier (1882).

[Wiltschko and Wiltschko 1996, Wiltschko and Wiltschko 2005]. This finding constitutes so far the main experimental proof of magnetoreception and magnetic compass orientation in birds. The possibility to study the orientation behavior of birds in the laboratory has allowed to accumulate a considerable experimental information on the dependence of the orientation ability on various external physical factors [Wiltschko and Wiltschko 1995, Wiltschko and Wiltschko 1996, Wiltschko *et al* 2005, Johnsen and Lohmann 2005, Wiltschko *et al* 2001, Wiltschko *et al* 2004, Mucheim *et al* 2002, Ritz *et al* 2004, Thalau *et al* 2005, Stapput *et al* 2006].

***2. The physiological and experimental basis for current magnetoreception models.***
Several models of magnetoreception have been proposed (for a review, see [Johnsen and Lohmann 2005]). Most of them, however, are not supported by any physiological or experimental evidence. For this reason, two hypotheses are now most popular: one involving magnetic minerals [Kirshvink and Gould 1981], and another one based on spin chemistry [Ritz *et al* 2000]. The latter, so-called radical-pair model will be discussed in the next Section.

The magnetic-mineral theory is largely based on universal presence of iron-containing compounds in living organisms. Recently, finding of a candidate magnetoreceptor structure in birds has been reported [Fleissner *et al* 2003, Fleissner *et al* 2007]. A complex structure attached to a nerve ending was found in the upper beak of the homing pigeon; remarkably, this structure contains large amounts of iron in various forms (mainly oxides $Fe_2O_3$ and $Fe_3O_4$ in different crystalline modifications, some of the crystals probably carrying macroscopic magnetic moments). In principle, the torque experienced by magnetic microcrystals in the external magnetic field can be detected by biological mechanoreceptors [Adair 2000]; however, no plausible explanation as to how exactly this specific structure might work has been so far proposed (see critical comments in [Winklhofer and Kirschvink 2008]).

Many experimental results indicate that a different mechanism of magnetoreception is responsible for determining the magnetic-field *direction*, i.e. for functioning of the bird's magnetic compass. It is believed that the compass receptor is situated in the bird's eye; this location is derived from the observed sensitivity of the bird's orientation ability to the intensity and spectral composition of the ambient light [Wiltschko *et al* 2001, Mucheim *et al* 2002, Wiltschko *et al* 2004], and from experiments with blindfolded birds. For instance, European Robins with covered right eye were shown to loose the ability to orient by the magnetic field, while those with covered left eye retained this ability [Wiltschko *et al* 2002]. Though this lateral asymmetry may be related to the structure of the bird's brain, the very fact that blocking vision of one or another eye disrupts orientation suggests that some structures in the eye are involved in magnetoreception. Accordingly, it was supposed that the magnetoreception is realized via a photochemical reaction sensitive to magnetic fields [Ritz *et al* 2000][2].

*3.The radical-pair model of magnetoreception.*

 The radical-pair model [Ritz *et al* 2000] assumes that the geomagnetic field changes the rate of radical-pair chemical reactions, affecting certain (so far unknown) receptors. Reactions of this type are well known in chemistry [Turro 1983, Salikhov *et al* 1984]. In such a reaction, an organic molecule is transferred into an excited state by absorption of a photon, and then splits in two radicals with spins of the two electrons forming a triplet state. The destiny of the molecule now depends on its spin state: if it remains a triplet, a chain of reactions, involving each of the two radicals separately, follows; if it transforms into a singlet, the radicals rapidly recombine. The triplet-to-singlet transition may result from the hyperfine interaction: one of the electrons flips its spin under influence of magnetic fields created by nuclear spins nearby. The probability of this process depends on the external magnetic field. Therefore, the yield of the photoinduced reaction may be, to a degree, controlled by the magnetic field. Calculations show that certain (though rather weak) sensitivity to magnetic fields of the order of the geomagnetic one (50 $\mu$T) may be expected [Adair 1999]. The sensitivity to the *direction* of the magnetic field is attributed to the anisotropy of the hyperfine interaction with respect to structural axes of the molecule: the molecules are assumed to be spatially fixed and oriented uniformly, in order to make the overall reaction yield a function of the field direction [Ritz *et al* 2000]. Recently, an argument in favor of this hypothesis was gained by biochemical methods: cryptochrome, a protein able to undergo a chain of photoinduced transformations affected by magnetic fields, was found in the birds' retina [Mouritsen *et al* 2004].

---

[2] This is a modification of an earlier proposal, not involving photoexcitation [Schulten *et al*, 1978].

### *4. Experiments on the influence of radio-frequency magnetic fields on magnetic orientation of birds.*

In an attempt to prove the radical pair model, a series of experiments was made that gave very remarkable results, which will be discussed in the rest of the paper. These are the experiments on the effect of weak oscillating magnetic fields on the work of the birds' magnetic compass [Ritz *et al* 2004, Thalau *et al* 2005, Stapput *et al* 2006, Wiltschko *et al* 2007]. In these experiments, the ability of caged birds to show the seasonally appropriate direction of movements was tested under application of alternating magnetic fields with amplitudes of the order or less than 0.01 of the constant geomagnetic field. The results may be summarized as follows:

1) A broadband "noise" magnetic field with the frequency range 0.1-10 MHz and average amplitude of 85 nT, applied at $24^o$ to the local geomagnetic field of 46 $\mu$T [Ritz *et al* 2004, Wiltschko *et al* 2007], disrupted the orientation of birds.

2) Orientation of birds was also disrupted by monochromatic oscillating fields of the amplitude of 485 nT in the frequency range, approximately, from 1 to 7 MHz (experiments at 1.315 [Thalau *et al* 2005, Stapput *et al* 2006], 2.63 [Stapput *et al* 2006] and 7MHz [Ritz *et al* 2004] have been so far reported). The oscillating field was applied vertically (i.e. at the angle of $24^o$ to the local geomagnetic field).

3) A higher sensitivity to the oscillating field was observed at 1.315 MHz: the orientation ability was disrupted by RF fields 10 times weaker than at other frequencies [Stapput *et al* 2006] [3].

4) Oscillating fields with the same amplitude, but applied parallel to the geomagnetic field, did not affect the orientation of birds, with the only exception for 1.315 MHz, at which frequency a marginally weak effect was observed [Ritz *et al* 2004, Thalau *et al* 2005].

In the original papers, these findings were interpreted as the electron paramagnetic resonance in the radicals undergoing the radical-pair reaction. Given the frequency range and strength of the ac magnetic field, it is indeed impossible to explain these effects by its influence on nuclear spins or motion of charges [Adair 1991, Adair 2000]. However, the analysis given in the next Section shows that these results are not explainable in terms of the electron spin resonance in organic radicals either.

---

[3] It is erroneously stated in [Thalau *et al* 2005, Stapput *et al* 2006] that 1.315MHz is the exact frequency of the paramagnetic resonance of unperturbed electron spins in the local geomagnetic field of 46 $\mu$T; in fact, this frequency is 1.289 MHz.

## 5. Experiments with the RF field: general phenomenology.

The following estimates are based on the general theory of the magnetic resonance, which can be found in many textbooks and monographs (see, for instance, [Abragam 1961, Abragam and Bliney 1970]).

We start with estimating the rate of electron spin transitions induced by the RF fields applied in the experiments. This can be most easily done for the experiment with the 0.1-10 MHz broadband field, because in this case we should not assume a specific resonance frequency (which is not known *a priori*). The general expression for the transition rate between quantum states $i$ and $j$ under a random magnetic field $B_1$ reads:

$$W_B = \frac{1}{\hbar^2} |\langle i|M_B|j\rangle|^2 \int_0^\infty \langle B_1(t)B_1(t-\tau)\rangle d\tau = \frac{1}{\hbar^2} |\langle i|M_B|j\rangle|^2 \langle B_1^2\rangle \tau_c \qquad (1)$$

where $\langle i|M_B|j\rangle$ is the matrix element of the projection of the electron magnetic moment on the direction of the RF field, and the correlation time of the random field is determined as $\tau_c = \frac{1}{\langle B_1^2\rangle} \int_0^\infty \langle B_1(t)B_1(t-\tau)\rangle d\tau \approx (2\pi\Delta f)^{-1}$, where $\Delta f$ is the bandwidth of the random field. Eq.(1) is valid if the transition frequency $\omega_{ij}$ satisfies the condition $\omega_{ij}\tau_c \ll 1$ (approximation of short correlation time). In terms of the bandwidth this condition reads $\omega_{ij} \ll 2\pi\Delta f$, which is indeed the case in the discussed experiments. Taking the transition matrix element equal to one of a free electron spin (this gives an upper estimate of $W_B$):

$$\langle i|M_B|j\rangle = \mu_B g B_1 \sin\alpha, \qquad (2)$$

we obtain, for $\Delta f = 10$ MHz, $\alpha = 24°$, and $B_1 = 85$ nT, the transition probability $W_B \approx 1 s^{-1}$. Here $\mu_B$ is the Bohr magneton, $\alpha$ is the angle between the RF and constant fields, and $g \approx 2$ is the electron g-factor,

Since the component of the RF field, parallel to the constant one, produces almost no effect, we should conclude that $i$ and $j$ are states with definite electron spin projections on the constant field $B_0$. For the RF field to induce a noticeable change in population of these states the transition probability should be of the same order as, or larger than, $T_1^{-1}$, where $T_1$ is the longitudinal spin relaxation time. Thus, we are coming to the following lower estimate of $T_1$:

$$T_1 > 1s \qquad (3)$$

Now we can use the experiments with monochromatic RF fields to estimate other characteristics of the spin system in question. With an exception for 1.315 MHz, the frequencies used in the experiment are not associated with any known transition, and it is very unlikely that they hit some exact resonance by chance. Therefore, we should consider these experiments as off-resonant, with the detuning of several MHz. The probability of such transitions is given by the quantum-mechanical perturbation theory [Landau and Lifshits 1973]:

$$W_{OR} = \left(\frac{\mu_B g B_1 \sin\alpha}{\hbar \Delta\omega}\right)^2 T_2^{-1} \qquad (4)$$

where $B_1$ is again the amplitude of the RF field, $\Delta\omega$ is the detuning, and $T_2$ is the dephasing time of the periodic perturbation. Since, as noted above, the transition goes between the states with definite projections of the electron angular momentum, the dephasing time $T_2$ can be interpreted in the usual for EPR manner as the transversal relaxation time.

Once again, the RF field inducing a noticeable change in population of these states means that the transition probability $W_{OR}$ is of the same order as, or larger than, $T_1^{-1}$. This condition results in the following relation of relaxation times:

$$\frac{T_2}{T_1} < \left(\frac{\mu_B g B_1 \sin\alpha}{\hbar \Delta\omega}\right)^2 \approx 10^{-6}, \qquad (5)$$

from which we readily find $T_2 \approx 10^{-6} s$.

Close to the exact resonance (that is, when $\Delta\omega << T_2^{-1}$), the transition probability equals

$$W_R = \frac{1}{2\hbar^2}(\mu_B g B_1 \sin\alpha)^2 T_2 \qquad (6)$$

It becomes larger than $T_1^{-1}$ under the condition

$$\frac{1}{2\hbar^2}(\mu_B g B_1 \sin\alpha)^2 > (T_1 T_2)^{-1} \qquad (7)$$

With the values of relaxation times that we have found, this condition is satisfied if $B_1 > 10 n\text{T}$, in qualitative agreement with the experimental result obtained for 1.315 MHz [Stapput *et al* 2006].

To conclude this Section, we have found that the observed disruption of the birds' orientation by the broadband noise magnetic field with the amplitude 85nT and frequency range of 0.1-10 MHz implies the longitudinal relaxation time $T_1$ of the electron spin system in question, approximately equal to 1s. This result does not depend on the specific energy spectrum of the system, provided the transition frequencies fall within the bandwidth of the RF field. The results of experiments with monochromatic RF fields at different frequencies can be qualitatively explained suggesting that the transversal relaxation time of the system, $T_2$, equals to $10^{-6}$ s.

*6. Restrictions imposed by the determined relaxation times on the spin system in question.*

The analysis of the experimental data, performed in the previous Section, reveals two unusual facts: a very long longitudinal relaxation time $T_1 \approx 1 s$ of the spin system in question, and a very large ratio of longitudinal and transversal relaxation times, $\frac{T_1}{T_2} \approx 10^6$.

The first fact immediately rejects the hypothesis that the observed disruption of birds' orientation is due to the effects of RF fields on radical pairs: typical lifetimes of such pairs are $10^{-9} - 10^{-7}$ s [Turro 1983, Adair 1999], and even the most optimistic estimates do not give lifetimes longer than $10^{-5}$ s [Cintolesi *et al* 2003]. Moreover, to the best of my knowledge, the time $T_1$ as long as one second was never observed in electronic spin systems of condensed matter.

The second finding, the large $T_1/T_2$, leads to even stronger conclusions. Large $T_1/T_2$ ratios in weak magnetic fields are typical for crystalline solids; they are never met in liquids or soft organic matter. The origins of this fact are the following. The longitudinal time $T_1$ is normally longer than the transversal time $T_2$ for two reasons. One is the suppression of relaxation of the spin component along the constant magnetic field $B$. But one can easily estimate that the constant field of $50\,\mu$T cannot suppress relaxation to the required degree. Indeed, the suppression factor reads

$$T_1/T_2 \approx 1 + \left(\mu_B g B \tau_c^* / \hbar\right)^2 \qquad (8)$$

where $\tau_c^*$ is the shortest of the two correlation times: one of effective random magnetic fields causing spin relaxation, and one of the electron spin itself. Without going into details of the spin relaxation mechanism, one can only say that $\tau_c \leq T_2$. By substituting this inequality into Eq.(8) and neglecting 1 in the right-hand side, one gets $T_1/T_2 \leq (\mu_B g B / \hbar)^2 T_2^2$. With $B=50\,\mu$T and $T_2=10^{-6}$, this gives $T_1/T_2 \leq 10^2$, which is far below the observed value $10^6$.

The second possible reason for a large $T_1/T_2$ is that the relaxation of the longitudinal spin component is accompanied by dissipation of the Zeeman energy $\mu_B g B$, which requires coupling of spin with the motion of atoms. This process is indeed strongly suppressed in solids with rigid crystal structure, but not in liquids and not in soft matter, where thermal agitation of molecules is much less restricted. For this reason, so large values of $T_1/T_2$ are met only in crystals, and even there they typically occur at low temperatures.

Therefore, we should conclude that the observed effects of RF magnetic fields on magnetic orientation of birds could not result from influence of these fields on electron spins in organic radicals.

## 7. External RF field vs internal random nuclear fields.

This conclusion is further confirmed by a comparison of the RF fields applied in the experiment with magnetic fields created by nuclear magnetic moments in the media surrounding the radical.

Each magnetic nucleus produces the field $\vec{B}_N = \frac{\mu_N}{Ir^3}\left(\vec{I} - 3\frac{(\vec{I}\cdot\vec{r})\vec{r}}{r^2}\right)$, where $\mu_N$ and $I$ are nuclear magnetic moment and spin respectively. The squared field is $B_N^2 = \frac{\mu_N^2}{I^2 r^6}\left(I^2 + 3\frac{(\vec{I}\cdot\vec{r})^2}{r^2}\right)$.

The mean squared field produced by all the nuclei inside a spherical layer with radius $r$ and thickness $dr$ at the center of the sphere is

$$\langle B_{Nr}^2 \rangle = \frac{\mu_N^2}{I^2 r^6}\left(I^2 + 3\frac{\langle(\vec{I}\cdot\vec{r})^2\rangle}{r^2}\right) \cdot 4\pi n_I r^2 dr = \qquad (9)$$

$$= \frac{\mu_N^2}{I^2 r^6}\left(I^2 + 3I^2\langle\cos^2\theta\rangle\right) \cdot 4\pi n_I r^2 dr = \frac{8\pi\mu_N^2}{r^4} n_I dr$$

where $\theta$ is the angle between $\vec{r}$ and $\vec{I}$, and $n_I$ is the concentration of magnetic nuclei. Integrating over $r$ from $R$ to infinity, we obtain the field produced by all the nuclei outside a "bubble" with radius $R$ at its center:

$$\langle B_{NR}^2 \rangle = \int_R^\infty \frac{8\pi\mu_N^2 n_I}{r^4} dr = \frac{8\pi\mu_N^2 n_I}{3r^3} \qquad (10)$$

For protons, $\mu_N = 1.41\cdot 10^{-26}$ J/T. In water, $n_I \approx 6\cdot 10^{22}$ cm$^{-3}$. We can use this value also as a rough estimate for $n_I$ in organic substances. So we get

$$\langle B_{NR}^2 \rangle^{1/2} = 30\mu T \cdot \sqrt{1/R^3}, \qquad (11)$$

where $R$ is in nm. Assuming $R=1$nm, we get $\langle B_{NR}^2 \rangle^{1/2} = 30\mu T$, i.e. the random nuclear field is 400 times stronger than the broadband RF field (85 $nT$) applied in the experiment! However, the correlation time of the nuclear field, $\tau_{cN}$, can be different from that of the RF field. In water, it is determined by molecular motion and is of the order of $10^{-11}$ s [Abragam 1961]. This means that typical frequencies of electron spin transitions are within the bandwidth $\Delta\nu_N$ of the random field created by water protons, and we can calculate the transition rate as $W_N = \dfrac{\gamma_e^2 \langle B_N^2 \rangle}{2\pi \cdot \Delta \nu_N} = \gamma_e^2 \langle B_N^2 \rangle \tau_{cN} \approx 3 \cdot 10^2 \, s^{-1}$, that is, more than 2 orders of magnitude faster than that induced by the broadband RF field.

If the radical is surrounded by soft matter (lipids or proteins), the correlation time of the nuclear field is longer because the molecular motion is much slower, while its amplitude is nearly the same as in water. As a result, the transition rate under random nuclear fields is in this case even faster.

Therefore, external RF fields applied in the discussed experiments exert much weaker effect upon electron spins in radicals than internal nuclear magnetic fields inevitably present in biological systems; under these conditions, external RF fields are not expected to affect radical pair reactions.

*8. Other possible origin of the effect: magnetic nanoparticles.*

If organic radicals cannot be responsible for the observed effect of RF fields on the bird's compass, what other object can? According to the results of Sections 6 and 7, it should be a piece of crystalline solid, which does not contain magnetic nuclei. In addition, it should be large enough to diminish penetration of nuclear magnetic fields from protons contained in the organic matter in which it is immersed. Using Eq.(11), it is easy to estimate that its radius should be not less than 20-30 nm. Finally, this object should carry an electron spin moment.

Seemingly, this is the portrait of an iron oxide magnetic nanocrystal!

However, spin relaxation times presently known for magnetic nanoparticles are far shorter than those estimated in Section 5. For example, in a recent work [Noginova *et al* 2007] superparamagnetic $\gamma - Fe_2O_3$ particles were studied by magnetic resonance in the external magnetic field of about 0.3T. The longitudinal relaxation time was measured to be just 10 ns, instead of 1s required.

On the other hand, experimental data on spin relaxation in magnetic nanocrystals are sparse, leaving some room for speculations. For instance, I am unaware of any such experiment performed in magnetic fields as weak as the geomagnetic field. Possibly, small Zeeman splitting together with a modification of the spectrum of lattice vibrations by size effects in the nanocrystal [Goupalov and Merkulov 2001] can suppress energy transfer from electron spins to the crystal lattice and, correspondingly, make $T_1$ longer. Still, gaining in $T_1$ by 8 orders of magnitude does not seem realistic. I would like to stress once again that the electron spin relaxation time of 1 s at the temperature of 310K is unheard of in condensed matter.

## 9. Conclusions

As our analysis has shown, the radical pair model is unable to explain the results of the experiments [Ritz *et al* 2004, Thalau *et al* 2005, Stapput *et al* 2006, Wiltschko *et al* 2007] on disruption of the bird magnetic compass by radio-frequency magnetic fields. The main reasons are that i) this would require longitudinal spin relaxation times of the order of 1s, which is much longer than the lifetime of the radical pair, and ii) internal stochastic magnetic fields produced by magnetic nuclei around the radicals exert much stronger effect on their spins than the external RF fields applied in the experiments. Moreover, the properties of the hypothetic object, that would give the required response to magnetic fields of amplitudes and frequencies used in those experiments, are so unusual, that they can hardly be attributed to any known physical system.

The experimental procedure of Refs.[ Ritz *et al* 2004, Thalau *et al* 2005, Stapput *et al* 2006, Wiltschko *et al* 2007] looks adequate for a biological experiment, and the conclusions would have been quite reasonable if they were purely biological. But it follows from these experiments that the compass magnetoreceptor of birds incorporates an unknown object with physical properties that have not been yet realized in physical laboratories. Therefore, the methodological requirements of experimental physics should be also applied to these experiments. And one of these requirements is that, if an experimental result contradicts the general theory, the experiment should be repeated with a special attention to all imaginable parasitic effects that could give the same result – until all of them are, one by one, excluded. Only then the new phenomenon can be believed existing. This was expressed in a concise form by Robert Adair [Adair 1999] : "Remarkable conclusions, which seem to violate well considered principles, require remarkably strong evidence". The Adair's maxim seems fully in place here.

If further research confirms the so far obtained results, they may lead to a breakthrough in the condensed-matter physics, where systems with long spin relaxation times are hunted for because of their envisaged technical applications ("spintronics" [Dyakonov 2004]). This would mean that Nature has by far outperformed the attempts of physicists and engineers to make practical use of long-lived spin states in condensed matter.

**Acknowledgements**

I would like to express my gratitude to the biologists who introduced me to the problem of animal magnetoreception, especially to Dr. D.Giunchi and Mr. D.Kishkinev. Special thanks are due to Profs. R. and W.Wiltschko for detailed discussions of the latest achievements in this field and explanation of their experimental procedure, and to Dr. J.Bojarinova for constant support and numerous discussions on various aspects of ornithology.